\documentclass[aps,superscriptaddress,twoside,a4paper,floatfix,nofootinbib]{revtex4}
\usepackage{hyperref,graphicx,amsmath,amssymb,color,times}

\newcommand{\bra}[1]{\langle #1|}
\newcommand{\ket}[1]{|#1\rangle}
\newcommand{\Exp}[1]{\langle#1\rangle}

\begin{document}
\title{Quantum Cheshire Cats}
\author{Yakir Aharonov} \affiliation{Tel Aviv University, School of Physics and Astronomy, Tel Aviv 69978, Israel}\affiliation{Schmid College of Science, Chapman University, 1 University Drive, Orange, CA 92866, USA}
\author{Sandu Popescu} \affiliation{H. H. Wills Physics Laboratory, University of Bristol$\text{,}$ Tyndall Avenue, Bristol, BS8 1TL, United Kingdom}
\author{Daniel Rohrlich} \affiliation{Physics Department, Ben Gurion University of the Negev, Beersheba, Israel}
\author{Paul Skrzypczyk}\affiliation{Department of Applied Mathematics and Theoretical Physics$\text{,}$ University of
Cambridge, Centre for Mathematical Sciences, Wilberforce Road, Cambridge CB3 0WA, United Kingdom}

\begin{abstract}
In this paper we present a quantum Cheshire Cat. In a pre- and post-selected experiment we find the Cat in one place, and its grin in another. The Cat is a photon, while the grin is its circular polarization.
\end{abstract}

\maketitle

\section{Introduction}
\begin{quote}
	\emph{\hspace{0.5cm}`All right,' said the Cat; and this time it vanished quite slowly, beginning with the end of the tail, and ending with the grin, which remained some time after the rest of it had gone.\\
	\indent\hspace{0.5cm}`Well! I've often seen a cat without a grin,' thought Alice, 'but a grin without a cat! It's the most curious thing I ever saw in my life!'}
\end{quote}

No wonder Alice is surprised. In real life, assuming that cats do indeed grin, the grin is a \emph{property} of the cat -- it makes no sense to think of a grin without a cat. And this goes for almost all physical properties. Polarization is a property of photons; it makes no sense to have polarization without a photon. Yet, as we will show here, in the curious way of quantum mechanics, photon polarization may exist where there is no photon at all. At least this is the story that quantum mechanics tells via measurements on a pre- and post-selected ensemble.
\section{Cheshire Cats}

In the following experiment, the ``cat" is a photon in two possible locations, $\ket{L}$ and $\ket{R}$.  The ``grin" corresponds to its circular polarization state.  The two basis states for circular polarization are $\ket{+}$ and $\ket{-}$.  In terms of horizontal and vertical linear polarization states $\ket{H}$ and $\ket{V}$, respectively, they are $\ket{+}= \left( \ket{H} +i\ket{V}\right)/\sqrt{2}$ and $\ket{-}=\left( \ket{H} -i\ket{V}\right)/\sqrt{2}$.

\begin{figure}[t]
	\includegraphics[width=0.25\columnwidth]{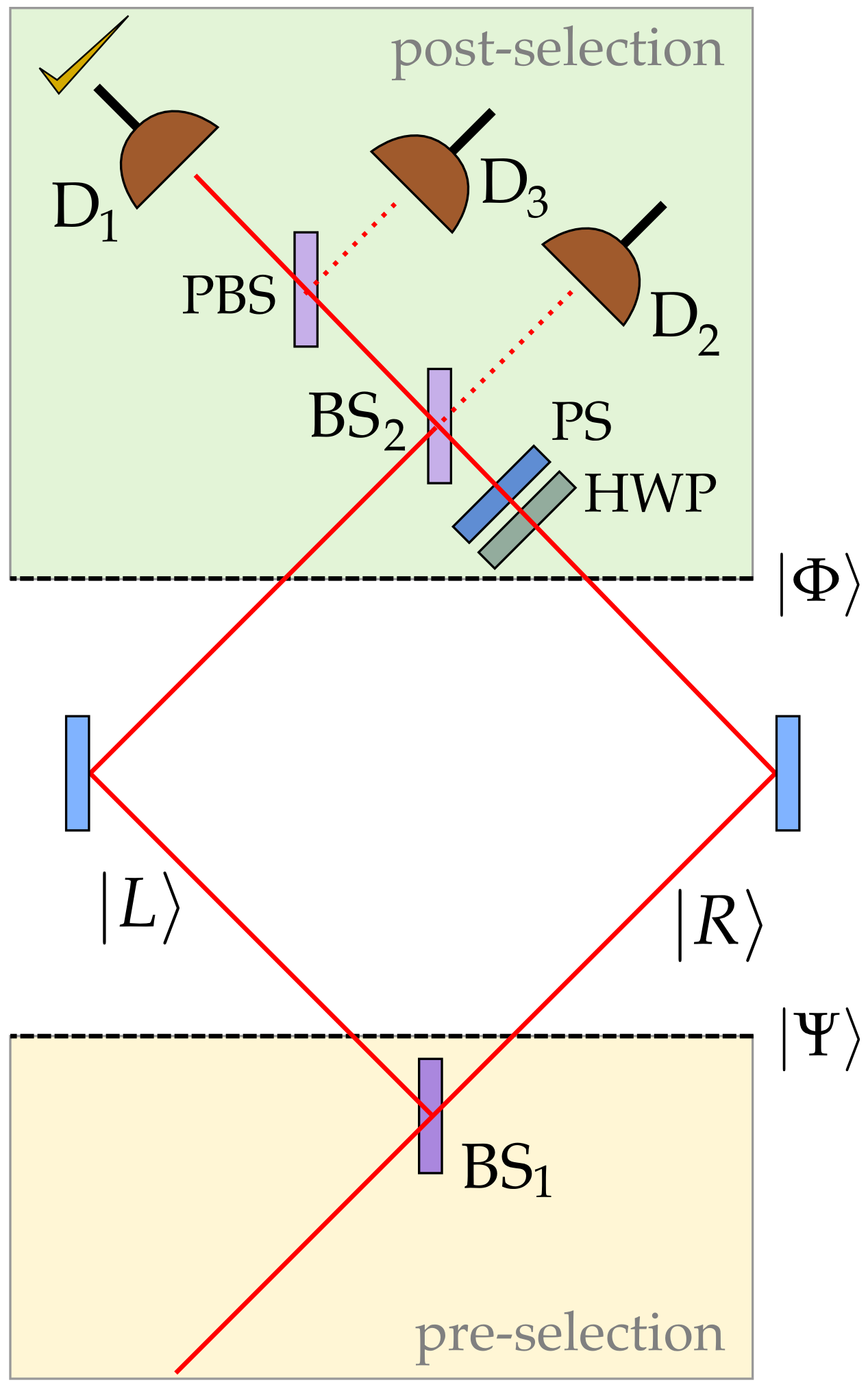} \caption{\label{f:setup} {\bf Schematic diagram of setup.} Various measuring devices will be inserted into the left and right arms of the interferometer, between the pre- and post-selection. }
\end{figure}

Suppose that the photon is initially prepared in a state $\ket{\Psi}$,
\begin{equation}
	\ket{\Psi} = \tfrac{1}{\sqrt{2}}(i\ket{L}+\ket{R})\ket{H},
\end{equation}
which is in a superposition of two locations $\ket{L}$ and $\ket{R}$ and horizontally polarized. A simple way to prepare such a state is to send a horizontally polarized photon towards a 50:50 beam splitter, as depicted in Fig.~\ref{f:setup}. The state after the beam splitter is $\ket{\Psi}$, with $\ket{L}$ now denoting the left arm and $\ket{R}$ the right arm; the reflected beam acquires a relative phase factor $i$.

We would like to post-select the state $\ket{\Phi}$,
\begin{equation}
	\ket{\Phi} =
 \tfrac{1}{\sqrt{2}}\big(\ket{L}\ket{H} + \ket{R}\ket{V}\big).
\end{equation}
In other words, we would like to perform a final measurement that gives the answer ``yes'' with certainty whenever the system is in the state $\ket{\Phi}$ and the answer ``no'', again with certainty, whenever the state is orthogonal to $\ket{\Phi}$.  We will then consider only those cases in which the answer ``yes'' is obtained.  Such a measurement can be experimentally realized in an optics setup, as depicted in Fig.~\ref{f:setup}.  The measuring device comprises a half-wave plate (HWP), a phase shifter (PS), a beam splitter (BS$_2$), a polarizing beam splitter (PBS) and three photon detectors (D$_i$). The HWP is chosen such that $\ket{H} \leftrightarrow \ket{V}$. The PS is chosen to add a phase factor $i$ on the beam, BS$_2$ is chosen such that if a photon in the state $(\ket{L}+i\ket{R})/\sqrt{2}$ impinges upon it, then it will certainly emerge from the left port (i.e. the detector $D_2$ will certainty \emph{not} click). The PBS is chosen such that $\ket{H}$ is transmitted and $\ket{V}$ is reflected. Given these choices, if the state immediately before the HWP (i.e. the state of the photon entering the measuring device) is $\ket{\Phi}$, then D$_1$ will click with certainty.  A photon in any state orthogonal to $\ket{\Phi}$ will end up either at detector $D_2$ or at $D_3$. We thus want to consider the experimental arrangement depicted in Fig.~\ref{f:setup}, which is nothing more than a modified Mach-Zehnder interferometer equilibrated such that in the absence of the HWP and PS, a photon entering BS$_1$ from the left will certainly emerge from BS$_2$ towards the right. 

We will focus only on those cases in which detector D$_1$ clicks. Inside the interferometer (i.e. in between the regions denoted by pre- and post-selection in Fig.~\ref{f:setup}), the photon is thus described by the \emph{pre-selected state} $\ket{\Psi}$ and \emph{post-selected state} $\ket{\Phi}$. It is the properties of the photon in these pre- and post-selected states that is the focus of this paper.

Let us first ask which way the photon went inside the interferometer. We will show that, given the pre- and post-selection, \emph{with certainty the photon went through the left arm}.  For suppose that we check the location of the photon by inserting photon detectors into the arms of the interferometer.  Let them be non-demolition detectors in the sense that they do not absorb the photon and do not alter its polarization. In mathematical terms, these detectors measure the projection operators $\Pi_L= \ket{L}\bra{L}$ and $\Pi_R=\ket{R}\bra{R}$.  Suppose first that we insert one such detector into the right arm. Is it possible to find the photon there? No, it is not.  If we find a photon there, then the state $\ket{\Psi'}$ after this measurement will be $\ket{\Psi'} = \ket{R}\ket{H}$, which is orthogonal to the post-selected state $\ket{\Phi} = (\ket{L}\ket{H} + \ket{R}\ket{V})/\sqrt{2}$. Hence the post-selection could not have succeeded in this case (i.e. detector D$_1$ could not have clicked). Thus the non-demolition measurement in the right arm never finds the photon there, indicating that the photon must have gone through the left arm. If instead we perform a non-demolition measurement in the left arm, given the post-selection, it will always indicate that the photon is there. We can even perform non-demolition measurements in both arms simultaneously, and they will always indicate that the photon was in the left arm. 
\emph{The Cat is therefore in the left arm. But can we find its grin elsewhere?}

Suppose instead of the measurements above, we place a polarization detector in the right arm. Since we know the photon is never in the right arm, surely no `grin' can ever be found there, and this detector should never click. Surprisingly however, the polarization detector in the right arm does click. We will discover that \emph{there is angular momentum in the right arm}.

Formally, a polarization detector in the right arm can be defined as
\begin{equation}
	\sigma_z^{(R)} = \Pi_R \sigma_z
\end{equation}
where
\begin{equation}
	\sigma_z = \ket{+}\bra{+}-\ket{-}\bra{-}
\end{equation}
The observable $\sigma_z^{(R)}$ has three eigenvalues, $+1$, $-1$ and $0$, corresponding to the eigenstates $\ket{R}\ket{+}$, $\ket{R}\ket{-}$ and the degenerate subspace spanned by $\ket{L}\ket{+}$ and $\ket{L}\ket{-}$, respectively. 

If a photon ends up at D$_1$ then the corresponding measurement of photon position $\Pi_R$ never finds that photon in the right arm.  Yet, surprisingly, a measurement of $\sigma_z^{(R)}$ may sometimes find angular momentum there. Indeed, the conditional probability of $\sigma_z^{(R)}$ yielding the result +1, given that the photon ends up at D$_1$, is non-zero. Similarly, there is a non-zero conditional probability that the measurement will find angular momentum -1 in the right arm. 


We seem to see what Alice saw---a grin without a cat! We know with certainty that the photon went through the left arm, yet we find angular momentum in the right arm. 

But could this conclusion really be right?  It is, ultimately, open to the following criticism. We never actually simultaneously measured the location and the angular momentum. Indeed, our conclusions above were reached by measuring location on some photons and angular momentum on others. The immediate implication is that all we have here is a paradox of counterfactual reasoning, in a class with other such paradoxes in quantum mechanics, e.g. \cite{Boh51,Har92}. That is, we have made statements about where the photon is, and about where the angular momentum is, that are paradoxical as long as we don't actually perform all the relevant measurements simultaneously. But let us see what actually happens if we try to measure the location and the angular momentum at the same time. 

Suppose that we simultaneously insert detectors for $\Pi_R$, $\Pi_L$ and $\sigma_z^{(R)}$.   (Since $\Pi_R$ and $\sigma_z^{(R)}$ commute, their ordering in the right arm does not matter.)  What we see now is that whenever $\sigma_z^{(R)}$ indicates net angular momentum, $\Pi_R$  yields the value 1, indicating that the photon in fact went through the right arm; whenever $\sigma_z^{(R)}$ does not indicate angular momentum, $\Pi_R$ yields the value 0, indicating that the photon went through the left arm.  The paradox thus evaporates.  This is the standard resolution of such counterfactual paradoxes in quantum mechanics: measurements disturb each other \footnote{Surprisingly, when measured on a pre- and post-selected ensemble, even commuting observables such as $\Pi_R$ and $\sigma_z^{(R)}$ may disturb each other.} therefore the conclusions drawn from separate measurements do not hold when measurements are performed simultaneously. Hence one is tempted to conclude that the paradox is nothing other than an optical illusion. In the next section, however, we will show that there really is a Cheshire Cat and it is not an optical illusion.  But doing so requires a subtler method.

\section{Weak measurements}

We have reached the central claim of this paper.  Ending the analysis with the resolution just presented, which is the common way of resolving quantum paradoxes, would be premature and would miss the essence.  As discussed above, the disturbance due to intermediate measurements is a  standard rationale for dismissing such paradoxes.  However, there is always a trade-off between disturbance and precision . That is, the disturbance due to measurements can be limited, at the price of accepting a certain level of imprecision (i.e. errors) in the measurement. It is then interesting to see what such limited-disturbance measurements –- which are performed simultaneously –- can tell about our paradox.  As we will now show, by adopting this strategy we regain the paradox that was prematurely lost.

We shall first present the specific scheme we have in mind to perform a limited-precision limited-disturbance measurement. This scheme is is very similar to those used in certain optical beam experiments presented in Refs. \cite{HolKwi08,DixStaJor09}, and can be performed with present-day technology.

The detectors measuring $\Pi_L,~ \Pi_R,~ \sigma_z^{(L)}$ and $\sigma_z^{(R)}$ would be realized by replacing the detector D$_1$ with a CCD camera, with the vertical and horizontal displacements of the beam serving as measurement pointers. For example, a flat glass sheet in the left arm, with its normal tilted at a small angle above the beam axis, displaces upwards a photon passing through it, by a small amount that we can define to be one unit $\delta$ of displacement. Then observing such an upward displacement of the beam in the CCD camera will indicate photons passing through the left arm. Similarly, a measurement of angular momentum could be an optical element producing a horizontal displacement of the beam in accordance with photon polarization.

The beam will have a characteristic cross-sectional width or ``waist" $\Delta$. The precision of the measurement and the degree to which it disturbs the photon depend upon the magnitude of the displacement $\delta$ relative to the width $\Delta$. When $\delta$ is much larger than $\Delta$, the measurement is precise; we can say with certainty, for a given photon, whether it is displaced or not.  At the same time the disturbance of the photon is large, because the location of the beam becomes entangled with what is measured. By contrast, $\Delta \gg \delta$ characterizes the so-called \emph{weak measurement} regime, or in other words the regime of limited disturbance. In this regime, any given photon does not reveal whether the beam has been displaced or not; but repeating the measurement $N$ times reduces the uncertainty in the beam displacement to approximately $\Delta/\sqrt{N}$.  Thus the displacement can be detected to any desired accuracy by repeating the measurement sufficiently many times.

In the context of pre-and post-selection, the above strategy is a specific implementation of a general measurement strategy known as \emph{weak measurements}, and the results they yield, the so called \emph{weak values} \cite{AhaAlbVai88,AhaVai08}, have already given new insight into many paradoxical situations in quantum mechanics \cite{AhaPopRoh93,AhaRoh05,AhaBotPop02}. 

In more detail, denoting by $A$ any operator measured as above, or in any weak measurement scheme, it is well known from standard results that the average shift of the pointer (or in the above the average shift of the beam) will be
\begin{equation}
	\Exp{A}_w = \frac{\bra{\phi}A\ket{\psi}}{\langle \phi | \psi \rangle},
\end{equation}
where $\Exp{A}_w$ is the \emph{weak value} of $A$, and where $\ket{\psi}$ is the pre-selected state and $\ket{\phi}$ is the post-selected state. Again, to repeat, the value $\Exp{A}_w$ is what the pointer of the measuring device indicates when $A$ is measured, with a measurement interaction that disturbs the measured system only weakly, on an ensemble of systems all pre-selected in the state $\ket{\psi}$ and post-selected in the state $\ket{\phi}$.  Moreover $\Exp{A}_w$ is the effective value of the observable $A$ for any system interacting with this ensemble, as long as the interaction is weak \cite{RohAha02}.

Let us now consider the story of our setup, as told by the weak values. The weak values are as follows:
\begin{equation}\label{e:Pi1}
	\Exp{\Pi_L}_w = \frac{\bra{\Phi}\Pi_L\ket{\Psi}}{\langle \Phi | \Psi \rangle} = 1
\end{equation}

\begin{equation}
	\Exp{\Pi_R}_w = \frac{\bra{\Phi}\Pi_R\ket{\Psi}}{\langle \Phi | \Psi \rangle} = 0
\end{equation}

\begin{equation}
	\Exp{\sigma_z^{(L)}}_w = \frac{\bra{\Phi}\sigma_z^{(L)}\ket{\Psi}}{\langle \Phi | \Psi \rangle} = 0
\end{equation}

\begin{equation}\label{e:sz2}
	\Exp{\sigma_z^{(R)}}_w = \frac{\bra{\Phi}\sigma_z^{(R)}\ket{\Psi}}{\langle \Phi | \Psi \rangle} = 1,
\end{equation}
where we have defined $\sigma_z^{(L)}$ for the left arm in analogy with $\sigma_z^{(R)}$ for the right arm. Thus the story as told by the weak values is that the photon is in the left arm (since $\Exp{\Pi_L}_w = 1$ and $\Exp{\Pi_R}_w = 0$) while the angular momentum is in the right arm (since $\Exp{\sigma_z^{ (L)}}_w = 0$ and $\Exp{\sigma_z^{(R)}}_w = 1$). 

The crucial point is that in principle all of these values apply simultaneously, since all of the weak measurements can be performed at the same time. In our specific scheme we can only measure any two at the same time, for example $\Exp{\sigma_z^{ (R)}}_w$ and $\Exp{\Pi_R}_w$, which indicate that there is a grin but no cat in the right arm. Alternatively we can measure any other pair and all results will be consistent with the paradox. We have finally found our Cheshire Cat.

\section{$N$-electron Cheshire Cat}
So far, we have considered an optical realization of the Cheshire Cat. The reason is that the proposed optical experiment can be implemented with current technology, as we hope it soon will be.  A drawback of this particular optical realization, however, is that it reveals the Cheshire Cat only as an average over many repetitions of the experiment. In this section we describe an alternative setup, one which is beyond the reach of current technology, but which reveals the Cheshire Cat in a stronger sense. In this setup the Cheshire Cat is seen only rarely---yet when it is seen, it is seen \emph{unambiguously and not as an average.} 

Consider $N$ (distinguishable) electrons, prepared in a superposition of locations $\ket{L}$ and $\ket{R}$ in two boxes.  All the electrons are polarized along the $x$-axis. The pre-selected state of the electrons is
\begin{equation}
	\ket{\Psi_N} = 2^{-N/2}\left([\ket{L}+\ket{R}]\ket{\uparrow_x}\right)^{\otimes N},
\end{equation}
Let us imagine that we can perform a measurement which allows us to post-select the electrons in the state
\begin{equation}
	\ket{\Phi_N} = 2^{-N/2} \left(\ket{L}\ket{\uparrow_x}+\ket{R}\ket{\downarrow_x}\right)^{\otimes N},
\end{equation}
The probability of post-selecting this state is $|\langle\Phi_N|\Psi_N\rangle|^2 = 4^{-N}$, which is exponentially small in the number of electrons. But when this post-selection succeeds it yields a Cheshire Cat that can be detected and measured with high precision. Indeed, let the ``Cat'' itself---its position---be defined by the mass of the electrons. It will be possible to perform a measurement that is both weak (i.e. does not appreciably disturb the pre- and post-selected states) and precise (will yield the number of electrons in each box up to an uncertainty of $\sqrt{N}$, which is insignificant relative to $N$ when $N$ is large) \cite{AhaVai08}. Such a measurement, e.g. by means of a gravitational probe, will find all $N$ electrons (up to uncertainty $\sqrt{N}$) residing in the left box.  Namely, the weak value of $\sum \Pi_L$ between the pre- and post-selected states $\ket{\Psi_N}$ and $\ket{\Phi_N}$, respectively, is
\begin{equation}
\left\langle  \sum^N \Pi_L\right\rangle_w =\sum^N \left\langle \ket{L}\bra{L}\right\rangle_w = N,
\end{equation}
while the weak value of $\sum \Pi_R$ is
\begin{equation}
\left\langle  \sum^N \Pi_R\right\rangle_w =\sum^N \left\langle \ket{R} \bra{R} \right\rangle_w = 0.
\end{equation}
However, the ``grin'', which here could be the magnetic field in the $z$-direction, will be equally measurable, weakly and precisely, by a suitable magnetic probe. The field will be found emanating from the right box, with field strength proportional to $N$, again with uncertainty of $\sqrt{N}$.  That is, $\langle \sum \sigma_z\Pi_R\rangle_w =N$ and $\langle \sum \sigma_z \Pi_L \rangle_w =0$. Crucially, in principle all of these measurements can again be made simultaneously. In keeping with a general analysis \cite{AhaVai08}, this (technologically impractical) weak value and the previous (optical) weak value are conceptually equivalent.

\section{Conclusions}
We have shown that Cheshire cats have a place in quantum mechanics -- physical properties can be disembodied from the objects they belong to in a pre- and post-selected experiment. Although here we have only presented one example in full detail, where a photon is disembodied from its polarisation, it should be clear that this effect is quite general -- we can separate, for example, the spin from the charge of an electron, or internal energy of an atom from the atom itself. Furthermore it is important to realise that it is not just pointers of well-prepared measuring devices that indicate that the properties are disembodied -- any external system which interacts weakly with the pre- and post-selected system will react accordingly \cite{RohAha02}. 

This therefore opens many intriguing questions, both conceptual and applied ones. First of all, how will an electron with disembodied charge and mass affect a nearby electron? In an atom with the internal energy disembodied from the mass, what will the resulting gravitational field look like? What sort of thermal equilibrium will be achieved by a system whose two degrees of freedom are separated? Furthermore, when considering more than two degrees of freedom, can we separate them all from each other? Can photons impart angular momentum to one object while their radiation pressure is felt by another object?

On the applied side, we may ask whether Cheshire cats have the potential to be useful in precision measurements, just as weak measurements have now shown themselves to be useful as a powerful amplification technique \cite{ResLunSte04,PryOBrWhi05,HolKwi08,DixStaJor09,YokYamKoa09,PalMalNgu10}. As an example, suppose that we wish to perform a measurement in which the magnetic moment plays the central role, whilst the charge causes unwanted disturbances. The question that arises is whether it might be possible to remove this disturbance, in a post-selected manner, by producing a Cheshire cat where the charge is confined to a region of the experiment far from the magnetic moment. We believe that such potential applications are an interesting possibility that deserve further investigation. 

{\bf Acknowledgements} YA acknowledges support from the Israel Science Foundation. SP Acknowledges support from the Templeton Foundation and the ERC. SP and PS acknowledge support from the European integrated project Q-ESSENCE.

\bibliographystyle{/Users/paulskrzypczyk/Library/texmf/bibtex/bst/bjp_thesis_sty}
\addcontentsline{toc}{chapter}{Bibliography}

\begin{thebibliography}{10}

\bibitem{Har92}
L. Hardy, {\it Phys. Rev. Lett.} {\bf 68},  2981  (1992).

\bibitem{Boh51}
N. Bohr, in {\it Albert Einstein: Philosopher--Scientist}, ed. Paul A. Schilpp (New York:  Tudor Pub. Co.), 1951, pp. 201-41.

\bibitem{HolKwi08}
O. Hosten and P.~G. Kwiat, {\it Science} {\bf 319},  787  (2008).

\bibitem{DixStaJor09}
P.~B. Dixon, D.~J. Starling, A.~N. Jordan, and J.~C. Howell, {\it Phys. Rev. Lett.}
  {\bf 102},  173601  (2009).

\bibitem{AhaAlbVai88}
Y. Aharonov, D.~Z. Albert, and L. Vaidman, {\it Phys. Rev. Lett.} {\bf 60},  1351
  (1988).

\bibitem{AhaVai08}
Y. Aharonov and L. Vaidman, {\it Lect. Notes Phys.} {\bf 734},  399  (2008).

\bibitem{AhaPopRoh93}
Y. Aharonov, S. Popescu, D. Rohrlich, and L. Vaidman, {\it Phys. Rev. A} {\bf 48},
4084  (1993).

\bibitem{AhaRoh05}
Y. Aharonov and D. Rohrlich, {\em Quantum Paradoxes: Quantum Theory for the
  Perplexed} (Weinheim; Wiley--VCH), 2005.

\bibitem{AhaBotPop02}
Y. Aharonov, A. Botero, S. Popescu, B. Reznik, and J. Tollaksen, {\it Phys. Lett. A}
  {\bf 301},  130  (2002).

\bibitem{RohAha02}
D. Rohrlich and Y. Aharonov, {\it Phys. Rev. A} {\bf 66}, 042102 (2002).

\bibitem{ResLunSte04}
J.~K. Resch, J.~S. Lundeen, and A.~M. Steinberg, {\it Phys. Lett. A} {\bf 324},  125
  (2004).

\bibitem{PryOBrWhi05}
G.~J. Pryde, J.~L. O'Brien, A.~G. White, T.~C. Ralph, and H.~M. Wiseman, {\it Phys.
  Rev. Lett.} {\bf 94},  220405  (2005).

\bibitem{YokYamKoa09}
K. Yokota, T. Yamamoto, M. Koaski, and N. Imoto, {\it New J. Phys.} {\bf 11},  033011
   (2009).

\bibitem{PalMalNgu10}
A. Palacios-Laloy, F. Mallet, F. Nguyen, P. Bertet, D. Vion, D. Esteve, and
  A.~N. Korotkov, {\it Nature Phys.} {\bf 6},  442  (2010).

\end{thebibliography}

\end{document}